# Pneumatically controlled non-magnetic, high-power, and low insertion loss RF switch

Chavalchart Herabut, Bryan Rangel Valle, Vikram D. Kodibagkar, and Sung-Min Sohn, Member, IEEE

*Abstract*— While positive-intrinsic-negative (PIN) diodes are commonly used in radio frequency (RF) circuits, their use often degrades the signal-to-noise ratio (SNR) due to high insertion loss and interference from additional biasing circuit components, which is critical for SNR-prioritized applications. This work presents the design of a novel pneumatically controlled switch, called AeroSwitch, which serves as an RF switch alternative to PIN diodes by significantly reducing lossy elements and additional biasing circuitry, with a focus on the applicability of AeroSwitch for magnetic resonance imaging (MRI) RF switch implantation. AeroSwitch was assessed against the PIN diode using bench testing, revealing a slightly reduced average insertion loss of 0.1 dB and an enhancement in average isolation by 15 dB. The switches are incorporated into a capacitor switch array within an L-matching network. This matching network connects to a loop RF coil, serving as a high-impedance load. The quality (Q)-factors were evaluated compared to a PIN diode capacitor switch array matching network configured identically. Overall, the AeroSwitch demonstrated an average 62% improved Q-factor compared to the PIN diode. These findings indicate that AeroSwitch offers significant advantages over the PIN diode due to its lower loss and higher isolation, potentially improving SNR. Additionally, the temperature of the AeroSwitch, while operating at 100 watts, remained consistently below 30°C, suggesting its potential as a self-temperature-regulated high-power switch. Since the AeroSwitch is non-magnetic, has high-power capability, low insertion loss, requires no electric biasing, and uses no conductive wires, this switching technique is significantly beneficial for MRI applications and medical related other RF applications.

## I. INTRODUCTION

Radiofrequency (RF) switches are essential components in various RF circuits and systems, enabling precise control of signal flow across multiple pathways. Key performance parameters include insertion loss, which quantifies the power lost when the switch is in the ON state, influencing the system's noise vulnerability. Isolation measures the effectiveness of the switch in preventing RF signal leakage during the OFF state. Leakage can introduce unwanted signals into circuits, potentially causing interference and leading to circuit damage if excessive. The insertion loss and RF leakage can significantly impact the signal-to-noise ratio (SNR). Switching speed is another crucial parameter, indicating how rapidly the switch can change states of activation.

One of the most commonly and widely used RF switching components is the positive-intrinsic-negative (PIN) diode as a single-pole single-through (SPST) switch. When a DC forward-biased voltage is applied across the PIN diode, the diode turns on and allows RF signals to pass through. When the diode is subjected to reverse bias, RF signals are blocked [1], [2]. This property allows the PIN diode to switch via DC biasing, making it advantageous for applications requiring electronic control. However, due to the relatively high resistance of the internal PIN diode's structure compared to conductors like copper, PIN diodes exhibit a high insertion loss, which can degrade signal quality due to power loss. PIN Diodes also require multiple low-impedance capacitors or DC blocks to prevent the DC biasing signals from entering the RF source or load. These components are typically positioned on the main signal line, further introducing loss. High-impedance inductors or RF chokes are also required to stop RF signals from entering the DC biasing circuits. However, chokes still produce some RF leakage, causing unwanted RF coupling. Additional filters are often needed to eliminate RF leakage. Both insertion loss and RF leakage are nonlinear with respect to frequency, biasing conditions, and RF power levels [3]. All these effects should be considered for SNR-prioritized applications, such as magnetic resonance imaging (MRI) [4], [5], [6].

Recently, micro-electromechanical systems (MEMs) switch designs have been introduced in MRI, offering lower insertion loss, higher linearity, and improved isolation with lower RF leakage [7], [8]. However, both PIN and MEMs switches require electrical driver circuitry and high-voltage DC biasing, delivered through long cables from outside the RF-shielded MRI room. This setup often introduces significant RF noise on MR images (e.g., zipper artifact), generates eddy currents, and causes high-power RF transmit coupling. These factors also introduce electrical hazards associated with all electric components for MRI and must comply with the international electrotechnical commission (IEC) standard 60601, ensuring that the RF coil (i.e., MRI antenna) surface temperature does not exceed 43°C during MRI scanning [9], [10]. Thus, the use of electrically driven RF switches in MRI remains challenging in minimizing their negative impact on image quality (i.e., SNR) and safety.

To overcome these limitations, this study presents the AeroSwitch, a pneumatically controlled RF switch that features low-loss and high isolation, which outperforms PIN diode

The authors are with the School of Biomedical and Health Systems Engineering, Arizona State University, Tempe, AZ 85287 USA (e-mail: smsohn@asu.edu).

switches. The design underwent a comparative evaluation with the PIN diode switch in a bench test. We assessed and compared insertion loss, isolation, switching time, and heat dissipation. Additionally, the switches were assembled into an array and integrated into an L-matching network functioning as a matching capacitor array for an RF coil. Performance was demonstrated by matching the impedance of an RF loop coil load to 50 Ω and measuring the impedance matching range against a similar PIN diode capacitor array matching network. The analysis of reflection and Q-factor highlights the effectiveness of the low-loss switch.

While pneumatic control systems for MRI have been reported with high MRI-compatibility, their application to RF circuits for coils is novel [11], [12]. Unlike fast RF switching, such as the Transmit/Receive switch, this work focuses on a potential switching-based RF coil optimization, such as frequency tuning and impedance matching, that occurs before imaging and has no strict timing constraints [13], [14], [15]. Potential applications could include RF ablation, hyperthermia, interventional systems with MRI, and other medical devices requiring RF circuit controls.

## II. METHODS

### A. AeroSwitch

The switch consists of a hollow housing with an opening at the top for connecting an air tube, allowing air pressure to enter. The bottom of the housing features a pair of copper legs, which are soldered to a circuit where a switch is required. These copper legs lead to a pair of copper contact points inside the housing in Fig. 1(a). A sliding block positioned inside the housing also has a pair of electrically connected contact points. The sliding block moves in response to changes in air pressure. When pressurized, the block is pushed downward, securely connecting the contact points between the bottom of the housing and the sliding block, allowing electrical current to flow and activating the switch. When the pressure is negatively depressurized (i.e., under vacuum), the sliding block moves upward, disconnecting from the contact points and deactivating turning off the switch in Fig. 1(b). The switch is carefully designed to reduce minimal electric resistance and minimize parasitic reaction, ensuring efficient optimal performance. All the structural components were fabricated using a resin 3D printer (Photon Mono M5s Pro, Anycubic, China) with an XY resolution of 16.8 x 24.8 mm and the electrically conductive parts were manually formed by folding a 0.2mm thin pure copper sheet. The housing has dimensions of 6 mm x 6 mm x 12 mm, with internal contact points measuring 2mm x 1.5 mm on both sides. The measured weight of the sliding block is 63.1 mg, and it functions as a plunger driven by airflow. A mini air pump (Ultra Silent Mini 12V DC Air Pump, Jadeshay, China) provides an airflow of 2~3.2 liters per minute (LPM), generating a vacuum with a pressure below 420 millimeters per mercury (mmHg), which is used to operate the AeroSwitch. To maximize airflow control and efficiency, the top opening of the housing is sealed as a chamber with a Venturi nozzle to increase airflow velocity, while small holes in the housing are positioned below the chamber to enhance the pressure difference between the interior and exterior of the chamber. The performance of the AeroSwitch was evaluated by comparing it with a PIN diode switch through the different experiments described below.

1) Single AeroSwitch Evaluation

First, the AeroSwitch was soldered onto a small PCB milled using substrate material (RO4003C, Rogers Corporation, USA) as shown in Fig. 2(a, c). Two SMA connectors are connected to each leg of the AeroSwitch. The back side of the PCB was ground filled for a 50 Ω transmission line. Similarly, a low-loss, high-power, non-magnetic PIN diode (MA4P7470F-1072T, MACOM Technology Solutions Holdings, Inc., USA) was soldered onto an identical PCB of the same size, including the bias circuit, as shown in Fig. 2(b). The DC feeds into the PIN diode through an RF choke of 330nH inductor (IM02EBR33K, Vishay Intertechnology Inc., USA) and returns to the ground through another choke as shown in Fig. 2(d). A pair of 560 pF capacitors (DKD1111C05, Passive Plus Inc., USA) was added to the main RF line to prevent DC signals from passing through the ports. Additionally, a decoupling capacitor was placed near the DC feed to ground any RF leakage.

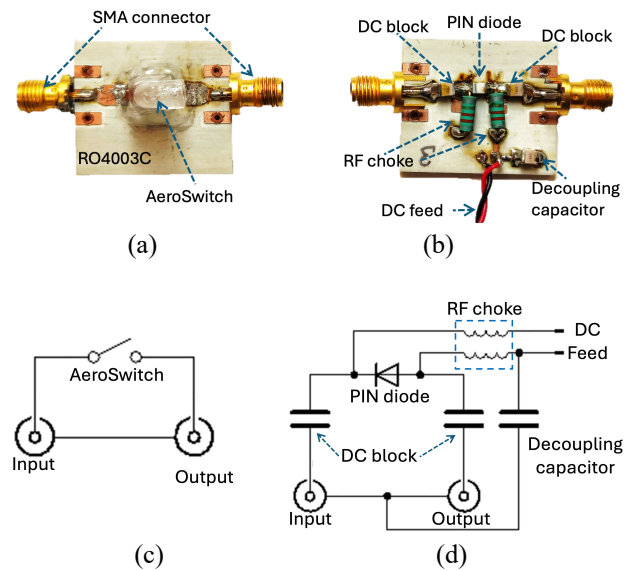

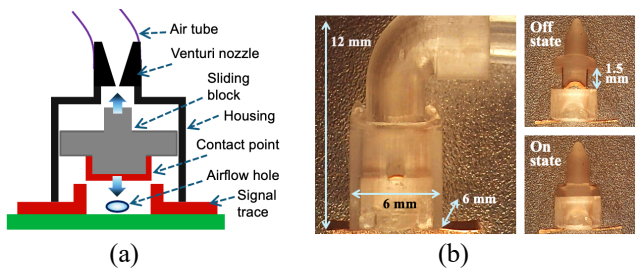

Fig. 1. (a) Diagram of the housing compartment of the AeroSwitch demonstrating all internal components and (b) photomicrograph of the fabricated AeroSwitch and its ON and OFF states.

Fig. 2. Photographs of (a) the fabricated single AeroSwitch and (b) PIN diode switch display component placement and routing. The circuit diagrams for (c) the AeroSwitch and (d) the PIN diode switch illustrates the electrical connections.

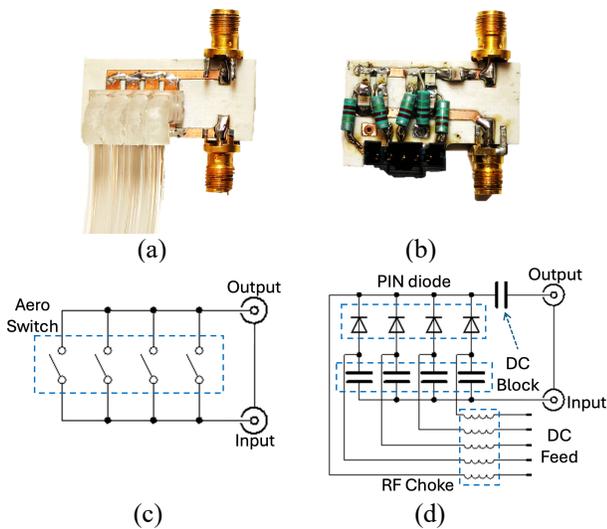

Fig.3. Photographs of (a) the fabricated AeroSwitch array and (b) PIN diode switch array. The circuit diagrams of (c) the AeroSwitch Array and (b) the PIN diode switch Array.

The transmission coefficient ($S_{21}$), representing insertion loss and isolation, was measured for both single switches in their on and off states across a frequency range of 100 MHz to 500 MHz using a two-port vector network analyzer (FieldFox N9923A, Keysight Technologies, USA).

To evaluate the switching times for the AeroSwitch in Fig. 2(a), a signal generator (SMC100A, Rohde & Schwarz, Germany) with a continuous wave of 300 MHz and 0 dBm power level was connected in the input port. An oscilloscope (InfiniVision MSOX3104T, Keysight Technologies, USA) was connected to the other port and measured control circuit latency time, RF pulses, and their rising and falling time. The timing for the AeroSwitch is broken down into two components: mechanical latency and switching time. Electrical and mechanical latency time is defined as the time difference between user input to RF signal response of the AeroSwitch. The switching time focuses specifically on the RF signal response change from 10% to 90% activation [16].

For high-power durability and temperature testing, a signal generator (SMC100A, Rohde & Schwarz, Germany) produced the RF source of a 500 MHz, 10 dBm signal, which was amplified to 100 W using a 40 dB gain amplifier (ZHL-100W-Gan+, Mini-Circuits, USA). The power line was connected to one of the single switch ports, and the other was terminated with a 50 Ω load (100-T-FN, Bird, USA). Temperature measurements were conducted using an infrared camera (TrueIR U5855A, Keysight Technologies, USA). Experiments lasted 10 minutes, with temperature readings taken every minute. Additionally, an infrared image was captured at the 10th minute to document the temperature distribution of both switching systems.

2) AeroSwitch Array Evaluation

To assess the functionality and effects of combining individual switches into a switching array, a four-AeroSwitches array was built. In Fig. 3(a, c), the switches are connected in parallel on a PCB with two ports without any components in series with the switches. This will ensure that the measured result will be the actual effect of the traces and the PCB geometry. Each switch is individually controlled by separate tubes linked to different channels of the air control station, enabling 16 possible switching combinations. A PIN diode switching array in Fig. 3(b) was constructed using a similar setup, incorporating the necessary biasing components with RF choke inductors and DC block capacitors with a similar value to the previous section. In this scenario, a DC block capacitor must be connected in series with each PIN diode, but it should be considered a very low impedance path because of its high capacitance. Fig. 3(d) illustrates the circuit diagram of all the components' connections. The $S_{21}$ parameters were measured across all 16 switching configurations for the AeroSwitch array and the PIN diode switch array.

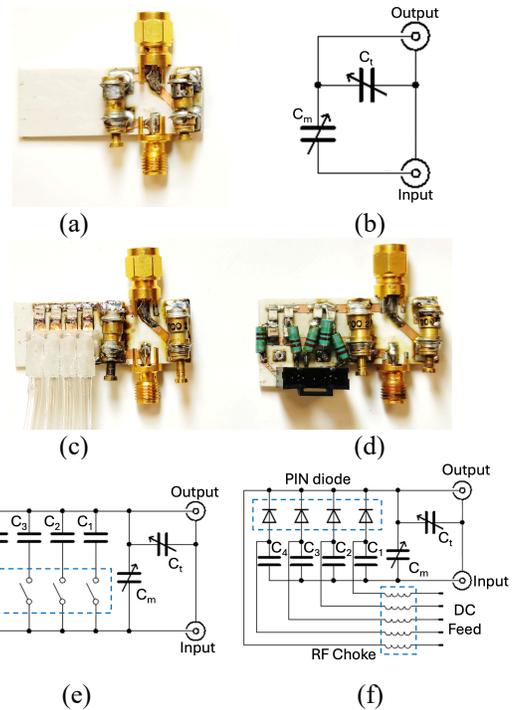

Fig. 4. L-Matching and tuning networks of (a) a basic circuit for MRI coils and (b) its circuit diagram. (c) AeroSwitch-based circuit, and (d) PIN Diode-based circuit with their corresponding circuit diagrams (e), and (f). And (g) experimental setup for comparing matching network performance.

3) L-Matching Circuit Testing

The switch arrays were then integrated into an L-type matching network. For the AeroSwitch matching network in Fig. 4(c, e), the circuit for the impedance matching consisted of a serial capacitor switch array in parallel with a trimmer capacitor (V9000, Knowles Voltronics, USA) with the value $C_m$ of approximately 6pF. The capacitor switch array was made with the AeroSwitch array, and each switch was serially connected to a lumped fixed-value capacitor (DKD1111C05, Passive Plus Inc., USA) with different values: $C_1$ = 1.0, $C_2$ = 1.2, $C_3$ = 1.5, and $C_4$ = 1.8pF. When the switches were activated, their respective capacitances were added to the capacitance of a matching circuit, resulting in different matching conditions. A shunt trimmer capacitor with the value $C_t$ approximately 10pF was added as an impedance tuning capacitor of the matching network. A PIN diode switching array was used to construct an L-type matching network in the same configuration to the AeroSwitch as shown in Fig. 4(d, f). DC block capacitors, RF choke inductors, and a DC feed connector were added to bias the PIN diodes, similar to the previous section. The matching networks have an input port for connecting to a source and an output port for connecting to a load.

A simple loop RF coil was designed to resonate at 298 MHz and used as a load. The coil was constructed using 80 mm diameter copper wire with a distributed capacitor of 2.7pF at the center of a loop. The phantom used was a human head-shaped container, filled with a solution that had a permittivity of 68 and a conductivity of 0.54 S/m, representing the average electrical properties of the human brain [17]. A basic L-type matching circuit in Fig. 4(a, b) without any capacitor switch array was also fabricated for control comparison. All three matching circuits were connected to the loop coil as a load, and the return loss ($S_{11}$) and input impedance were measured for both when the coil was loaded and unloaded with the phantom across all 16 switching configurations using a realistic MRI experimental setup, as shown in Fig. 4(g). The Q-factor was extracted from the $S_{11}$ data using (1) where $f_c$ is the central frequency and $df$ is the -3 dB bandwidth of a resonance [18]. The Q-factor determines the effect of the loss of the matching network

$$Q = f_c/df \qquad (1)$$

*B. Air Control System*

The air control system in Fig. 5 was designed to supply adequate airflow through tubing to the AeroSwitches mounted on a switches array, enabling their independent activation and deactivation via pressure and vacuum, respectively. Each AeroSwitch is connected to a pair of directional mini air pumps within the same tube network. One pump supplies positive pressure to the network through its output connector, while the other generates negative pressure by connecting to the network via its suction connector. Since the pumps cannot maintain pressure on their own, each pump is paired with a normally closed solenoid valve (KL04000, Hangzhou Beduan Ecommerce Co., China), positioned between the pump and the network. As the valves and pumps must activate simultaneously, each pump-valve pair is driven by a single MOSFET trigger switch drive module (ANMBEST, Wuhan, China). The MOSFET modules are electronically controlled by a microcontroller (Arduino Nano, Arduino, Italy), which reads user input from three-state switches configured in a pull-down setup. The system keeps all valves and pumps turned off when the user input switch is neutral. When switched to the second state, the positive pressure pump is activated, and its valve opens, allowing pressure to fill the tube and activate the AeroSwitch. In the third state, the suction pump engages, opening its valve to create a vacuum in the tube, thereby deactivating the AeroSwitch. Because the pump and vacuum unit is designed to control a single AeroSwitch, four identical units were replicated to enable independent control of each switch.

III. RESULT AND DISCUSSION

With the single switch setups in Fig. 2, the results of switching time, insertion loss, isolation, and the temperature of the AeroSwitch and PIN diode are illustrated and discussed as follows.

The switching time was measured from the user input signal at the air station to the activation of the conductive plunger using non-overlapped control signals to avoid a situation where pressure and vacuum occur simultaneously. In Fig. 6(a), the transition times for both rising ($T_{LR}$) and falling ($T_{LF}$) latency were approximately 40 ms. The switching time was measured within a range of ~250 μs for both ON ($T_R$) and OFF ($T_F$) states as shown in Fig. 6(b). Fig. 8(c) shows the occasionally unstable switching caused by the bouncing of the contact when operated at excessively high pressure, even though it occurs inside the switching time range. In comparison, PIN diodes exhibit latency due to an electrical mechanism rather than mechanical properties and an average switching time of less than 200 ns [19]. While the AeroSwitch has slower switching times compared to PIN diodes, it is not a critical issue in applications that do not require nanosecond-range fast operation, such as control circuits for optimizing RF conditions. Furthermore, the design can be optimized for greater reliability by refining its components through improved manufacturing techniques and design enhancements of the internal contact structure. Since air pressure levels, the distance of air tubes from the air control station, and the moving distance range of a sliding block are all related to the entire performance including temporal variations, these factors should be considered for an optimized AeroSwitch system.

The measured average insertion loss of an AeroSwitch and a PIN diode from 100 MHz to 500 MHz in Fig. 7(b) is -0.15 dB and

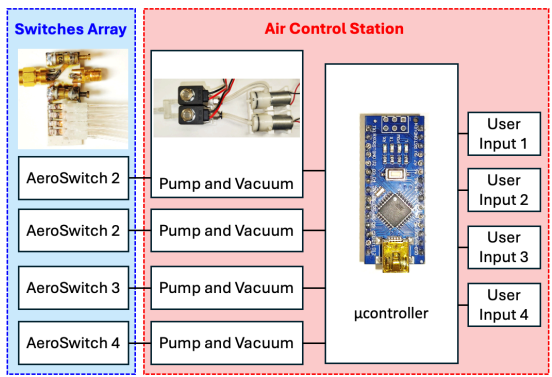

Fig. 5 Diagram of the air control system.

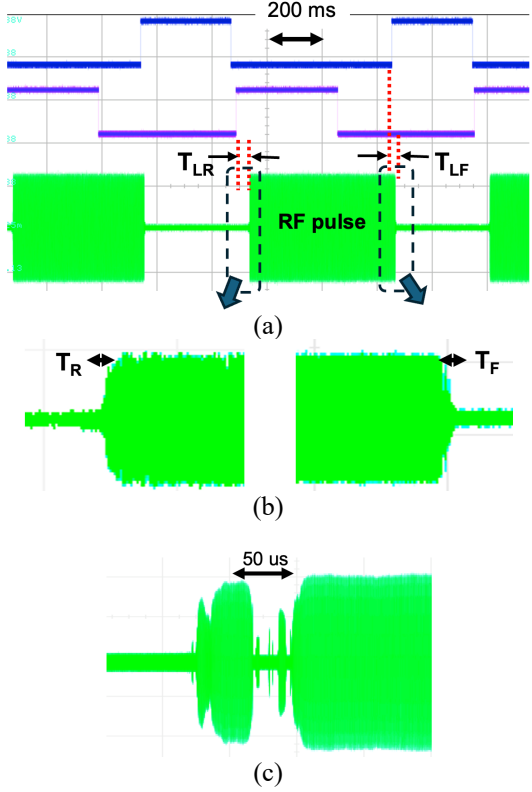

(a)

(b)

(c)

Fig. 6. Measured oscilloscope signals of (a) activation of the OFF state (top) and ON state (center) with electrical and mechanical latency timing of $T_{LF}$ and $T_{LR}$, as well as RF pulse (bottom), (b) switching time of the rising ($T_R$) and falling ($T_F$) time, and (c) potential unstable switching behavior.

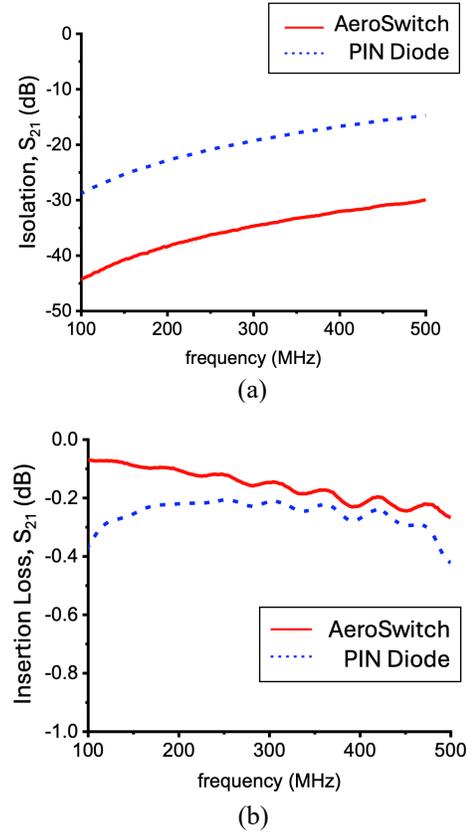

(a)

(b)

Fig. 7. Measured $S_{21}$ of both AeroSwitch and PIN diode switch using VNA for (a) isolation while the switches are deactivated and (b) insertion loss when the switches are activated.

-0.25 dB, respectively. Therefore, the AeroSwitch has an insertion loss of 0.1 dB higher than that of the PIN diode switch. The AeroSwitch has an average isolation of -35 dB, while the PIN diode exhibits -20 dB. This indicates that the AeroSwitch's isolation is 15 dB lower than that of the PIN diode switch, as illustrated in Fig. 7(a). These results suggest that more power can flow through the AeroSwitch during the ON state, while less power can leak through the AeroSwitch during the OFF state. This is expected because the AeroSwitch is simply a conductor line, which should have very low resistance compared to the intrinsic resistance of the PIN diode. The measured data reflects the general frequency-dependent behavior of RF switches, driven by their intrinsic characteristics. As discussed earlier in the switching time analysis, the insertion loss and isolation are also determined by various factors. For example, a longer moving distance increases switching time and results in greater isolation.

The temperature measurement in Fig. 8. shows the PIN diode in the ON state reaches 29.7°C, only slightly higher than the AeroSwitch at 26.9°C. However, when the switch remains in the OFF state for 10 minutes, the temperature of the PIN diode rises significantly to 40.5°C due to the high RF power dissipation. In contrast, the AeroSwitch exhibits only a minor temperature increase to 27.9°C, demonstrating superior thermal management. This difference suggests that the AeroSwitch's airflow-assisted cooling effectively regulates its temperature, reducing the risk of overheating. Improved thermal stability, as observed in the

AeroSwitch, is particularly beneficial for RF applications such as MRI, where maintaining consistent performance over extended periods, typically at least 10 minutes per image acquisition, is crucial for minimizing signal distortion and ensuring compliance with safety regulations under high RF power levels, typically around 1 kilowatt to induce MR signals. The SNR in MRI can be defined as

$$SNR \propto B_1 / \sqrt{R_s T_s + R_C T_C} \qquad (2)$$

Where $B_1$ is the induced RF magnetic field, $R_S$ and $R_C$ are the electrical resistance of the sample and coil, and $T_S$ and $T_R$ are their temperatures, respectively [20]. The SNR is affected by the positive feedback loop between resistances ($R_S$ and $R_C$) and temperatures ($T_S$ and $T_C$). Specifically, an increase in temperature leads to a rise in resistances (e.g., for copper, resistance increases by approximately 0.4% per 1°C temperature increase, and higher resistance, in turn, causes further temperature elevation. Alternatively, lowering $T_C$ is essential for suppressing thermal noise, and improving the SNR [21].

Next, the experiment analyzing the switching array setups depicted in Fig. 3 was conducted. The $S_{21}$ measurement of the AeroSwitch array shown in Fig. 9(a) reveals an isolation level below -15 dB across the entire frequency range when all switches remain deactivated, corresponding to the 0000 configuration. In contrast, an insertion loss exceeding -1.1 dB is noted for all other combinations (0001 to 1111). Fig. 9(b) presents the $S_{21}$ of the PIN diode switch array, exhibiting isolation below -12 dB, which is slightly underperformed than that of the AeroSwitch array. Upon activation, the average insertion loss across all configurations is around -3 dB, lower than that of the AeroSwitch array. The AeroSwitch demonstrates superior isolation and insertion loss performance compared to the PIN diode, which is consistent with findings from the single switch experiment. However, the overall isolation level observed in this experiment was significantly higher than that of the signal switch experiment. This might be attributed to the capacitive coupling between the PCB's parallel traces rather than the switches themselves. Both the AeroSwitch and PIN diode switch arrays feature parallel PCB traces. Thus, this effect is consistent in both setups without affecting the comparison of switch performance. The variations in insertion loss across switching configurations are minimal for both setups, indicating minimal influence from combining the switch into an array. Throughout the frequency range, the AeroSwitch exhibits a very linear $S_{21}$ response, while the PIN diode shows fluctuations with frequency. Furthermore, the broad bandwidth of testing complicates the maintenance of consistent behavior of the PIN diodes, including the bias circuit components, which change with frequency and may introduce additional attenuation at specific frequencies due to their parasitic characteristics.

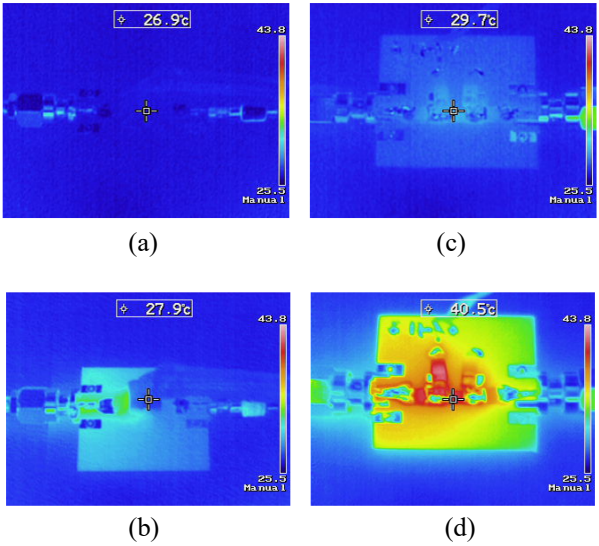

Fig. 8. Temperature comparison of the AeroSwitch and PIN diode in different states: (a) AeroSwitch in the ON state, (b) AeroSwitch in the OFF state, (c) PIN diode in the ON state, and (d) PIN diode in the OFF state.

After the performance of the AeroSwitch array was validated, it was integrated into a matching network for a more practical experiment in Fig. 4(g) to evaluate the matching and tuning capability and the Q-factor.

Without the matching network connected to the coil, the input impedance of the coil was measured to be at 32+j259.2 Ω at 300 MHz, which is highly mismatched from a 50 Ω system. With the matching network, the Smith charts display the measured input impedance by looking into the matching circuit for all setups across all 16 switching positions for both loaded and unloaded conditions in Fig 10. As the switching positions transition from 0000 to 1111, the input impedance changes due to the addition of matching capacitance, which modifies the matching conditions. This alteration causes the impedance points to move along an arc on the Smith chart. The trimmer capacitors were tuned so that the middle of the switching position matched the load to 50 Ω, which is the center of the Smith chart. The cluster size of impedance points shows a close similarity between the PIN diode and the AeroSwitch matching network setups. The cluster is more spread out under the unloaded conditions in Fig. 10(a) compared to the loaded conditions in Fig. 10(b) because the coil is more sensitive to matching conditions when the loss from the load is present. This demonstrates the functionality of the matching network integrated with the proposed AeroSwitch.

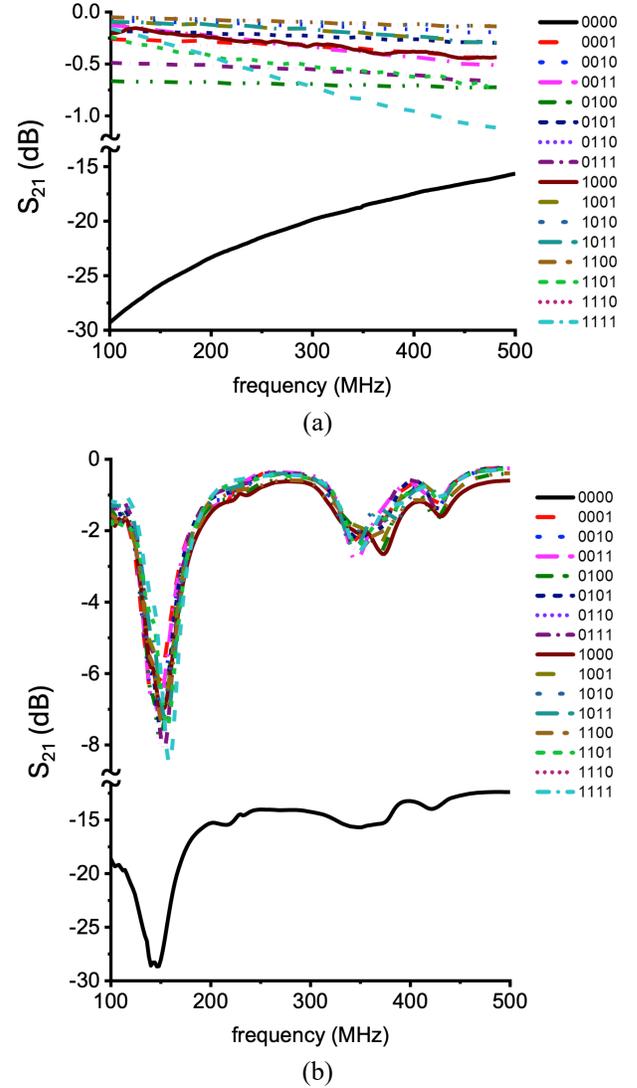

Fig. 9. Measured $S_{21}$ for all 16 switching combinations of (a) AeroSwitch array and (b) PIN diode switch array.

Table I. All 16 switching configurations are displayed as 0000 to 1111. Each digit corresponds to the activation of the four capacitors, where 0 indicates OFF and 1 indicates ON. Since the capacitors are connected in parallel, the summed capacitance of the activated capacitors, including the trimmer,

is shown in the second column. Q-factors extracted from $S_{11}$ of the standard, PIN diode, and AeroSwitch for both loaded and unloaded conditions are listed.

TABLE I
Q-FACTOR FOR ALL SWITCHING CONFIGURATIONS

| Switching Configuration | C(pF) | Q-factor | | | | | |
|---|---|---|---|---|---|---|---|
| | | Standard unloaded | Standard loaded | PIN Diode unloaded | PIN Diode loaded | AeroSwitch unloaded | AeroSwitch loaded |
| 0000 | 6 | 146.5 | 30.8 | 90.5 | 27.3 | 146.7 | 47.1 |
| 0001 | 7 | | | 69.1 | 26.1 | 146.6 | 35.6 |
| 0010 | 7.2 | | | 69.1 | 25.5 | 117.3 | 42.0 |
| 0011 | 8.2 | | | 53.3 | 24.4 | 90.1 | 33.6 |
| 0100 | 7.5 | | | 65.2 | 26.1 | 117.1 | 32.6 |
| 0101 | 8.5 | | | 48.8 | 23.4 | 97.5 | 27.9 |
| 0110 | 8.7 | | | 58.6 | 24.4 | 83.5 | 30.9 |
| 0111 | 9.7 | | | 46.8 | 23.0 | 89.8 | 26.6 |
| 1000 | 7.8 | | | 61.7 | 25.0 | 97.5 | 31.7 |
| 1001 | 8.8 | | | 48.8 | 23.0 | 89.9 | 27.9 |
| 1010 | 9 | | | 58.6 | 23.9 | 73.0 | 27.9 |
| 1011 | 10 | | | 45.0 | 22.1 | 83.4 | 26.0 |
| 1100 | 9.3 | | | 53.2 | 23.0 | 89.8 | 26.6 |
| 1101 | 10.3 | | | 39.0 | 21.3 | 89.8 | 23.9 |
| 1110 | 10.5 | | | 50.9 | 22.5 | 89.8 | 25.5 |
| 1111 | 11.5 | | | 40.3 | 21.6 | 89.8 | 24.4 |

The Q-factor of the coil across all setups was derived from the measured $S_{11}$ data. The Q-factors for all setups are summarized in Table I. The Q-factor of the standard setup in unloaded conditions exceeds 100, indicating that the coil and other components exhibit minimal loss and are optimal for this experiment, as referenced in [10]. In unloaded conditions, the phantom's loss lowers the Q-factor to 30, and the Q-ratio, defined as Q-ratio = Q-unloaded/Q-loaded = 3.33, illustrates that the coil is highly coupled with the phantom, which is beneficial for increasing signal sensitivity, leading to SNR improvement.

In comparing the AeroSwitch and PIN diode setups under no load, the AeroSwitch has an average Q-factor of 99.5, outperforming the PIN diode's Q-factor of 56.2. When loaded with the phantom, both setups see a reduced Q-factor. However, AeroSwitch's average Q-factor drops to 30.7, which is still higher than the 23.9 for PIN diodes. The combined average Q-factor for the AeroSwitch in both loaded and unloaded states is 65.1, compared to 40 for the PIN diode. This indicates that the AeroSwitch exceeds the PIN diode switch's average Q-factor by 62%. It's important to mention that when multiple switches are ON, the Q-factor significantly declines due to the increased capacitance observed in both setups under loading conditions.

The Q-factor comparisons between the two setups suggest that the PIN diode setup experiences higher losses than the AeroSwitch setup, which aligns with findings from two prior experiments. This could be attributed to the AeroSwitch's lower resistance than the intrinsic resistance of the PIN diode.

From the three experiments conducted, the mechanical switching tests demonstrated lower switching speeds compared to the PIN diode's electrical switching capabilities, making the AeroSwitch more suitable for applications requiring lower switching speeds. The temperature testing revealed superior thermal regulation compared to the PIN diode, indicating its potential for high-power applications. Across all experiments, the AeroSwitch exhibited lower insertion loss and higher isolation, making it well-suited for efficient power transmission. The Q-factor calculations in the matching network experiments highlighted differences in loss mechanisms, while Smith chart analysis provided insights into impedance variations, demonstrating the practical application of the AeroSwitch for impedance matching. These findings offer a comprehensive comparison, showcasing AeroSwitch's advantages in minimizing power dissipation and improving efficiency for specific RF switching applications. One potential application is preclinical MRI, where SNR is a critical factor. The AeroSwitch's ability to reduce losses makes it a promising candidate for such systems.

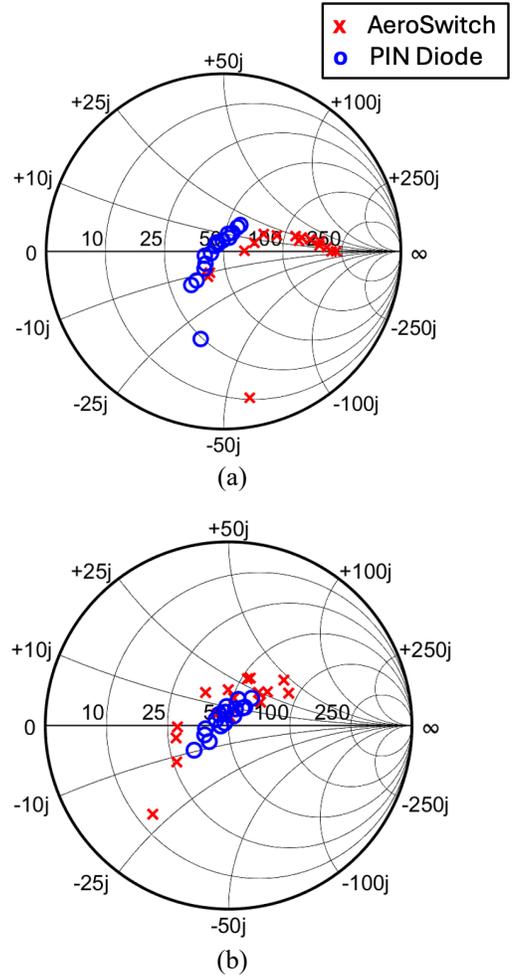

Fig. 10. The Smith chart displays input impedance across all switching combinations: (a) the impedance without a load and (b) with a load using both AeroSwitch and PIN diode.

Despite its strong performance in most aspects, the AeroSwitch could be further improved through advanced fabrication techniques. Optimizing its design may reduce its size, conserving space while also minimizing potential signal attenuation caused by parasitic effects.

IV. CONCLUSION

In this work, we designed a novel pneumatically controlled RF switch mechanism, the AeroSwitch, engineered with non-magnetic property and minimal electrical components.

Evaluation results comparing the AeroSwitch to a PIN diode demonstrated lower insertion loss and higher isolation. Additionally, temperature testing revealed the potential to be a self-temperature-regulated switch, particularly for high-power RF circuits. The practical application of a capacitor switching array for impedance matching an RF coil was successfully demonstrated, showing a high Q-factor potentially improving the RF coil's sensitivity in MRI applications. While the switch has switching speed and reliability limitations, further research on precise fabrication processes using micromachining tools and improved air pressure control could enhance reliability and reduce switching time.